\documentstyle[12pt]{article}
\newcommand{\es}{\\[2mm]}



\def\Lp{\displaystyle{\biggl(}}
\def\Rp{\displaystyle{\biggr)}}

\newcommand{\lp}{\left(}\newcommand{\rp}{\right)}

\newcommand{\lac}{\left\{}\newcommand{\rac}{\right\}}

\renewcommand{\a}{\alpha}
\renewcommand{\b}{\beta}
\renewcommand{\d}{\delta}
\newcommand{\e}{\varepsilon}
\newcommand{\f}{\phi}
\newcommand{\g}{\gamma}

\newcommand{\m}{\mu}
\newcommand{\n}{\nu}

\renewcommand{\o}{\omega}

\newcommand{\s}{\sigma}

\newcommand{\DD}{{\cal D}}

\newcommand{\FF}{{\cal F}}

\newcommand{\SS}{{\cal S}}

\newcommand{\ZZ}{{\cal Z}}

\newcommand{\complex}{{\kern .1em {\raise .47ex
\hbox {$\scriptscriptstyle |$}}
    \kern -.4em {\rm C}}}
\newcommand{\real}{{{\rm I} \kern -.19em {\rm R}}}
\newcommand{\rational}{{\kern .1em {\raise .47ex
\hbox{$\scripscriptstyle |$}}
    \kern -.35em {\rm Q}}}
\renewcommand{\natural}{{\vrule height 1.6ex width
.05em depth 0ex \kern -.35em {\rm N}}}
\newcommand{\dint}{\displaystyle{\int}}

\newcommand{\drit}{\dint d^2 \! x \, }

\newcommand{\cb}{{\bar c}}

\newcommand{\pa}{\partial}

\newcommand{\dfrac}[2]{{\displaystyle{\frac{#1}{#2}}}}

\newcommand{\sla}{\raise.15ex\hbox{$/$}\kern -.57em}

\newcommand{\twiddle}{\lower.9ex\rlap{$\kern -.1em\scriptstyle\sim$}}

\newcommand{\equ}[1]{(\ref{#1})}

\newcommand{\eq}{\begin{equation}}
\newcommand{\eqn}[1]{\label{#1}\end{equation}}
\newcommand{\eea}{\end{eqnarray}}
\newcommand{\eqa}{\begin{eqnarray}}
\newcommand{\eqan}[1]{\label{#1}\end{eqnarray}}
\newcommand{\ba}{\begin{array}}
\newcommand{\ea}{\end{array}}
\newcommand{\eqac}{\begin{equation}\begin{array}{rcl}}
\newcommand{\eqacn}[1]{\end{array}\label{#1}\end{equation}}

\newcommand{\at}{{\~a}}

\newcommand{\ooo}{{\'o}}

\newcommand{\iii}{{\'\i}}

\begin{document}


{\ }

\vspace{3cm}
\centerline{\LARGE Vector supersymmetry of Chern-Simons theory}\vspace{2mm}
\centerline{\LARGE at finite temperature}

\vspace{1cm}

\centerline{\bf {\large D.G.G. Sasaki}}
\vspace{2mm}
\centerline{\it C.B.P.F}
\centerline{\it Centro Brasileiro de Pesquisas F{\iii}sicas,}
\centerline{\it Rua Xavier Sigaud 150, 22290-180 Urca}
\centerline{\it Rio de Janeiro, Brazil}
\vspace{3mm}
\centerline{\bf {\large S.P. Sorella, V.E.R.Lemes}}
\vspace{2mm}
\centerline{{\it UERJ}}
\centerline{{\it Universidade do Estado do Rio de Janeiro}}
\centerline{{\it Departamento de F{\iii}sica Te{\ooo}rica}}
\centerline{{\it Instituto de F{\iii}sica, UERJ}}
\centerline{{\it Rua S{\at}o Francisco Xavier, 524}}
\centerline{{\it 20550-013 Maracan{\at}, Rio de Janeiro, Brazil}}
\vspace{4mm}


\centerline{{\normalsize {\bf REF. UERJ/DFT-03/99}} }

\vspace{4mm}
\vspace{10mm}

\centerline{\Large{\bf Abstract}}\vspace{2mm}
\noindent
The existence of the vector supersymmetry is analysed within the context of the finite temperature Chern-Simons theory. 
\newpage

\section{Introduction}

Since many years the topological three-dimensional Chern-Simons \cite{w,bbrt} theory is the source of continuous and renewed interests, with many applications going from pure field theory to condensed matter physics. The Chern-Simons gauge model has been the first example of a topological field theory of the Schwarz type, allowing for the computation of several topological invariants in knots theory \cite{w}. It is a remarkable fact that these computations can be performed within the standard perturbation theory \cite{eg}. Moreover, the Chern-Simons provides an example of a fully ultraviolet finite field theory, with vanishing  $\beta$-function and field anomalous dimensions \cite{uf}.  This  feature relies on the existence of an additional global invariance of the Chern-Simons action which shows up only after the introduction of the gauge fixing  and of the corresponding Faddeev-Popov ghost term. This further symmetry is known as {\it vector supersymmetry} \cite{dgs,bbrt} since its generators carry a Lorentz index and, together with the BRST symmetry, give rise to a supersymmetric algebra of the Wess-Zumino type. 
It worth mentioning  that the nonzero temperature version of the Chern-Simons action is also available \cite{cst} and turns out to play an important role in the applications of three-dimensional gauge theories to finite temperature effects. Therefore, it seems naturally to ask ourselves if the vector supersymmetry is still present in the case of a nonzero temperature. This is the aim of the present letter. In particular, we shall be able to show that this question can be answered in the affirmative. In this sense, the fully quantized Chern-Simons action can be considered as an example of a superymmetric field theory at finite temperature.
The paper is organized as follows. In Sect.2 we present the finite temperature Chern-Simons action and we analyse the existence of the aforementioned supersymmetry. Sect.3 will be devoted to the study of some consequences and to the conclusion. 

\section{Finite temperature Chern-Simons action}

In order to analyse the properties of the Chern-Simons action at finite temperature let us first recall the supersymmetric structure of the zero temperature case. Adopting the Landau gauge, for the fully quantized Chern-Simons action we have
 \eq
\SS = \int_{{\cal R}^3} d^{3}x \lp \dfrac{1}{4}\e^{\m\n\s} ( A^{a}_{\m}F^{a}_{\m\n}
- \dfrac{1}{3} f_{abc}A^{a}_{\m}A^{b}_{\n}A^{c}_{\s} ) +
b^{a}\pa^{\m}A^{a}_{\m} - \cb^{a}\pa_{\n}(D^{\n}c)^{a} \rp .
\eqn{CS3}
Expression \equ{CS3} is left invariant by the following nilpotent BRST transformations
\eq\ba{ll}
 s A^{a}_{\m} = (D_{\m}c)^{a} \, ,
& s c^{a}  = -\dfrac{1}{2}f^{abc}c^{b}c^{c}  \,  \es
 s \cb^{a} = b^{a} \, ,
& s b^{a}  = 0 \, .
\ea\eqn{brst}
In addition, the action \equ{CS3} is known \cite{dgs,bbrt} to possess a further rigid invariance whose generators $\d_{\m}$ carry a vector index, {\it i.e.}
\eq\ba{ll}
\d_{\m}c = A_{\m} \, , & \d_{\m}\cb = 0 \, \es
\d_{\m}b = \pa_{\m}\cb \, , & \d_{\m}A_{\n} = - \e_{\m\n\s}\pa^{\s}\cb , \es
\ea\eqn{susy3}
and, together with the BRST transformations, obey the following relations
\eq\ba{ll}
s^{2} = 0 \, , \qquad \lac \d_{\m}, \d_{\n} \rac = 0 \, \ , \es 
\lac \d_{\m}, s \rac = \pa_{\m} + {\ }{\ }{\rm eqs. \hspace{.1cm} of \hspace{.1cm} motion} \ ,
\ea\eqn{algebra}
which, closing on-shell on the space-time translations, give rise to a supersymmetric algebra of the Wess-Zumino type.

Concerning now the nonzero temperature case, for the quantized Chern-Simons action in the imaginary time formalism \cite{cst}, we obtain
\eq
\SS_{T} = \int_{0}^{\b} d\tau \drit \lp \dfrac{1}{4}\e^{\m\n\s} ( A^{a}_{\m}F^{a}_{\m\n}
- \dfrac{1}{3} f_{abc}A^{a}_{\m}A^{b}_{\n}A^{c}_{\s} ) +
b\pa^{\m}A_{\m} - \cb^{a}\pa_{\n}(D^{\n}c)^{a} \rp ,
\eqn{act}
where $\beta$ stands for the inverse of the temperature $T$. As is well known, all fields $\eta=(A,c,\cb,b)$ are required to obey {\it periodic} boundary conditions along the compactified  direction $\tau$ \cite{k}, namely 
\eq\ba{ll}
\eta(x,\tau) = {\displaystyle\sum}_{n=-\infty}^{\infty} \eta^{n}(x)e^{-i\o_{n}\tau}\, \ ,  \\[3mm]
\eta^{n}(x) = \dfrac{1}{\b}\dint_{0}^{\b} d\tau \eta(x,\tau)e^{i\o_{n}\tau} \, \ ,
\ea\eqn{tauexp}
where the $\o_{n}$ are the so-called Matsubara frequencies \cite{k}
\eq
\o_{n} = \dfrac{2\pi n}{\b}  \ .
\eqn{mats}
We emphasize here that the ghost fields $c,\cb$, although being anticommuting variables, have to be periodic in $\tau$. As we shall see in the following, this  property will be crucial for the existence of a supersymmetric structure at nonzero temperature.
In order to write down the finite temperature Chern-Simons action in terms of the Matsubara modes $\eta^{n}$, we identify the $\tau$-direction with the $x^3$ variable and we introduce the following useful two-dimensional notation
\eq\ba{ll}
 A_{\m}^{n} = (A_{\a}^n, \phi^n) \ , \qquad \a,\b,\g=1,2 \ , \es
\e^{3\m\n} = \e^{\a\b}\, \ , \qquad 
\e^{\a\b}\e_{\a\g} = \d^{\b}_{\g}  \ .
\ea\eqn{not}
Thus, for the action we obtain
\eq 
\SS_{T} = \SS_{inv} + \SS_{gf} \ ,
\eqn{b-act}
where 
\eq\ba{ll}
\SS_{inv} =\b \sum_{n} \drit    \Lp \dfrac{1}{2}\e^{\a\b} ( \f^{a n}F^{a \, -n}_{\a\b})
-\dfrac{i}{2}\o_{n}\e^{\a\b}A^{a \, n}_{\a}A^{a \, -n}_{\b} + i
\o_{n}b^{a\, n}\f^{a \, -n} \Rp \,  \es 
\SS_{gf} = \b \sum_{n} \drit  {\Lp} b^{a
\, n}\pa^{\a}A^{a \, -n}_{\a} +
 \pa_{\a}\cb^{a\, n} {\left( \pa^{\a}c^{a}
+ f^{abc}\sum_{l} A^{b \, l}_{\a}c^{-c(n+l)} \right)}  \Rp \,
\es
\hspace{.7cm} + \b \sum_{n} \drit  \Lp \cb^{a\, n}\o_{n}^{2}c^{a -n} + i\o_{n}\cb^{a\,
n}\sum_{l}f^{abc}\f^{b\,l}c^{-c(n+l)}
\Rp \, \\ \es
F^{a \, n}_{\a\b} = \pa_{\a}A^{a\,n}_{\b}-\pa_{\b}A^{a \,
n}_{\a}+f^{abc}\sum_{l}A^{b\,l}_{\a}A^{c\, (n-l)}_{\b} \ .
\ea\eqn{euclidesn}
In terms of the Matsubara modes, the BRST transformations \equ{brst} read
\eq\ba{ll}
 s A^{a\, n}_{\a} = \pa_{\a}c^{a\, n}
+ f^{abc}\sum_{l} A^{b \, l}_{\a}c^{c(n-l)}  \, ,
\es
 s\f^{a\, n} = -i\o_{n}c^{a\, n}+f^{abc}\sum_{l}\f^{b\, l}c^{c\, (n-l)} \,  \es
 s c^{a}  = -\dfrac{1}{2}f^{abc}\sum_{l}c^{b\, l}c^{c\, (n-l)} \, ,
\es
 s \cb^{a\, n} = b^{a\, n} \, ,
\es
s b^{a\, n}  = 0 \ .
\ea\eqn{brscomT}
Moreover, it can be checked that the nonzero temperature action \equ{b-act} is left invariant by the further following rigid transformations $\d_{\a},\d$, namely
\eq\ba{ll}
\d_{\a}A_{\b}^{a\, n} = i \o_{n}\e_{\a\b}\cb^{a\, n} \ , \es
\d_{\a}\f^{a\, n} = \e_{\a\b}\pa^{\b}\cb^{a\, n} \ , \es
\d_{\a}c^{a\, n} = A_{\a}^{a\, n} \ , \es
\d_{\a}b^{a\, n} =\pa_{\a}\cb^{a\,n} \ ,  \es 
\d_{\a}\cb^{a\, n} = 0 \ , 
\ea\eqn{b-susy1}
and
\eq\ba{ll}
\d A_{\a}^{a\, n} = -\e_{\a\b}\pa^{\b}\cb^{a\, n} \ , \es
\d c^{a\, n}= \f^{a\, n} \ , \es
\d \f^{a\, n} = 0 \ ,  \es
\d b^{a\, n}= -i\o_n \cb^{a\, n} \ , \es 
\d\cb^{a\, n} = 0 \ .
\ea\eqn{b-susy2}
The generators $\d_{\a},\d$ give rise, together with the BRST operator $s$, to the following algebraic relations
\eq\ba{ll}
\lac \d , s \rac \eta^n = -i\o_{n} \eta^n  + {\ }{\ }{\rm eqs. \hspace{.1cm} of \hspace{.1cm} motion} \ , \es
\lac \d_{\a} , s \rac \eta^n = \pa_{\a}\eta^n  + {\ }{\ }{\rm eqs. \hspace{.1cm} of \hspace{.1cm} motion} \ , \es
\lac \d_{\a},\d_{\b} \rac = 0 \ ,  \es
\d^{2}=0 \ .
\ea\eqn{b-algebra}
We see therefore that the supersymmetric structure \equ{algebra} of the zero temperature Chern-Simons persists also in the case of a nonvanishing temperature. In particular, it is easily recognized that the operator $\d$ of eqs.\equ{b-susy2} corresponds to the generator $\d_{\tau}$ of eqs.\equ{susy3} along the compactified direction $\tau=x^3$. It is also worth underlining here that the existence of a nonzero temperature supersymmetric algebra relies on the periodic boundary conditions required for the Faddeev-Popov ghosts $c, \cb$. As is well known, this property follows from the gauge invariance of the nonzero temperature action $\SS_{T}$. Moreover, the supersymmetry turns out to be crucial in order to ensure that no physical excitations show up in the nonzero temperature case, as it will be discussed in the next section. In other words, the nonzero temperature Chern-Simons action remains a topological theory, with no local physical degrees of freedom.

\section{Conclusion}

It has been already underlined that in the zero temperature case the existence of the vector supersymmetry is deeply related to the topological nature of the Chern-Simons term. We recall in fact that the supersymmetry shows up only after the introduction of the ghost fields. As a consequence, it follows that the contributions coming from the propagating components of the gauge field are exactly compensated by those corresponding to the ghosts, resulting in the well known ultraviolet finiteness of the theory. This means that the are no local physical degrees of freedom, {\it i.e.} that the theory is topological. 
The existence of a supersymmetric structure in the case of nonzero temperature 
suggests a similar behaviour for the  finite temperature version of the Chern-Simons. This fact can be easily confirmed in the  abelian case by showing that the partition function turns out to be  independent from the temperature, implying the vanishing of all relevant thermodynamic quantities.
Let us compute in fact  the partition function for the abelian Maxwell-Chern-Simons action
\eq
\ZZ = e^{-\b\FF} = \int {\DD A} {\DD c} {\DD b} {\DD \cb} e^{-\SS_{MCS}} \ ,
\eqn{z}
with 
\eq
\SS_{MCS} = \int_{0}^{\b} d\tau\drit \lac - \dfrac{g}{4}F_{\m\n}F^{\m\n} +
i\dfrac{1}{2}\e^{\m\s\n}A_{\m}\pa_{\s}A_{\n} + b\pa^{\m}A_{\m} - \cb
\pa^{\m}\pa_{\m}c \rac .
\eqn{M+CS}
We have introduced a constant $g$ in order to take into account the Maxwell term. Of course, the pure Chern-Simons contribution is recovered in the limit $g \rightarrow 0$. For the free energy $\FF$ we obtain the following result
\eq
\FF = -g L^2 \int \dfrac{d^{2}p}{4\pi^{2}}\lac  \sqrt{\vec{p}^{2} + 1/g^2} +
 \dfrac{2}{\b} ln(1 - e^{-\b\sqrt{\vec{p}^{2} + 1/g^2}}) \rac \ ,
\eqn{energia livre}
where $L^2$ stands for the two-dimensional area. Obviously, expression \equ{energia livre} does not depend from $\b$ in the limit $g \rightarrow 0$. Again, there is a complete compensation between the ghost and the gauge sectors, as expected from the existence of the supersymmetry. 
The analysis of the ultraviolet finiteness of the nonabelian finite temperature case as well as the computation of the vacuum expectation value of Polyakov loops are under investigation. 

\vspace{2cm}
\noindent
{\large\bf{Acknowledgements}}
\newline
The Conselho Nacional de Desenvolvimento Cient{\iii}fico e Tecnol{\ooo}gico CNPq-Brazil, the Funda{\c c}{\at}o de Amparo a{\`a} Pesquisa do Estado do Rio de Janeiro (Faperj) and the SR2-UERJ are acknowledged for the financial support.

\end{document}